\documentclass[12pt]{article}
\usepackage{pic02}
\usepackage{hyperref} 
\usepackage{url}
\usepackage{graphicx}

\usepackage{amsmath}
\usepackage{latexsym}
\usepackage{xspace}
\usepackage{relsize}

\newcommand{\gev}{\ensuremath{\mathrm{\,Ge\kern -0.1em V}}\xspace}

\def\babar{\mbox{\slshape B\kern-0.1em{\smaller A}\kern-0.1em
    B\kern-0.1em{\smaller A\kern-0.2em R}}\xspace}

\def\pb {\ensuremath{\rm \,pb}\xspace}

\def\invfb   {\ensuremath{\mbox{\,fb}^{-1}}\xspace}

\begin{document}

\title{\bf Prospects for R Measurement at \babar using Radiative Return}
\author{Nicolas Berger \\
{\em Stanford Linear Accelerator Center, 
2575 Sand Hill Road, Menlo Park, USA}}

\maketitle

\baselineskip=14.5pt
\begin{abstract}
A precise measurement of the ratio of hadron to muon production
in $e^+e^-$ collisions, denoted by $R(s)$, is a necessary input for the interpretation
of current precision electroweak measurements. We describe a method for measuring $R(s)$ with the
\babar detector at the PEP-II B-factory using initial-state radiation events.
\end{abstract}

\baselineskip=17pt

\section{Introduction}

In recent years, precision electroweak measurements have probed some of the fundamental parameters
of the Standard Model. Following the successful prediction of the mass of top quark,
the precise measurement of electroweak parameters can  now put significant constraints on 
the Higgs boson mass. Similarly, the unprecedented precision of the BNL measurement of the muon anomalous 
magnetic moment ($g-2$)~\cite{bnl} could also offer sensitivity to beyond-the-Standard-Model 
physics.

However, interpreting those results requires precise theoretical predictions of Standard Model contributions.
At present the leading source of uncertainty in these calculations arises from 
hadronic corrections, which cannot be computed from first principles. For the Higgs mass, this corresponds
mainly to hadronic contributions to the running of $\alpha_{QED}$, and  hadronic corrections to the QED vertex
for $g-2$.
These contributions can be related through dispersion integrals to the cross-section
for $e^+e^- \to \mbox{hadrons}$, or equivalently to $R(s)$, defined as its ratio 
with the Born cross-section for  $e^+e^- \to \mu^+ \mu^-$. For the running of $\alpha_{QED}$, 
and the muon magnetic moment, we have respectively
\begin{equation}
\Delta \alpha_{QED}^{Had}(M_Z^2) \propto \int_{4 m_{\pi}^2}^{\infty}{as \frac{R(s)}{s(s-M_Z^2)}} 
\hspace{0.5cm} \mbox{and} \hspace{0.5cm}
\Delta a_{\mu}^{Had} \propto \int_{4 m_{\pi}^2}^{\infty}{ds \frac{R(s)K(s)}{s}}
\end{equation}
where $K(s)$ is the QED kernel function.
A precise measurements of $R(s)$ over a wide energy range would have a significant 
influence on the determination of both these quantities.
It would also shed some light on recently discovered discrepancies between hadronic 
data from $e^+e^-$ collisions and $\tau$ decays, which have significant impact on the
interpretation of the $g-2$ measurement.~\footnote{See Andreas Hoecker's contribution in these proceedings.}
At present $R(s)$ can be reliably calculated above  $10 \gev$ using perturbative QCD. 
It would therefore be particularly useful to provide
a precise measurement for energies below $10 \gev$.

\section{The radiative return method}

There has recently been a renewed interest in the possibility that 
the the measurement of $R(s)$ could be performed at fixed-energy colliders by using radiative
return to lower energies~\cite{kuehn1, rodrigo2}. 
An especially attractive possibility would be to use the large event rates provided by the B-factories, 
where the fiducial cross-section is estimated to be of the order of
$40 \pb$ for events with a hadronic system invariant mass less than $7 \gev$.
, yielding about $3.6$ million fiducial events in the current \babar dataset of $90 \invfb$.

It would in particular be possible to perform an inclusive analysis, which would rely only 
on the identification of the tagged initial-state radiation (ISR) photon. A highly
efficient hadronic selection would be performed, with minimal sensitivity to detector efficiency
and event shape modeling.
Alternatively, an exclusive analysis could be performed, in which each possible hadronic decay channel
is analysed separately. This would be especially advantageous for energies near to the $\rho$ mass, which
provide the largest contribution to the $g-2$ integral,
since the measurement of $R(s)$ through
$\sigma(e^+e^- \to \pi^+ \pi^- \gamma)/\sigma(e^+e^- \to \mu^+ \mu^- \gamma)$ benefits from the cancellation 
of radiative and efficiency corrections.
The inclusive method would present the disadvantage that the photon energy resolution gives a 
large uncertainty on the effective center-of-mass energy $s'$ in the deep ISR regime. This would however 
not be a significant problem in the case
of the running of  $\alpha_{QED}$, for which the dispersion integral has little sensitivity to
uncertainties on $s'$. 

\section{Comparison with other methods}

Using the radiative return offers several advantages over energy-scan methods. The entire energy range --- for
\babar, from the $\rho$ peak to about $7 \gev$ --- can be covered in a single measurement.
 With the currently available datasets, the event rates in each energy region are expected 
to be competitive with other experiments:
the latest BES measurement of $R(s)$~\cite{bes}, in the energy range $2-5 \gev$ is based on
about $120,000$ events, while the current \babar dataset should include about $950,000$ events in the same interval.
The higher center-of-mass energy of the collision also leads to a different event geometry, with a collimated
hadronic system recoiling at high momentum against the ISR photon. By selecting tagged photons well within the 
detector we can ensure high fiducial acceptance for the hadronic system.
In addition, the boost imparted to the hadron system produces particles with 
higher transverse momentum, thus reducing the 
kinematic bias in the event selection due to limited $p_T$ coverage.
The high energy of the photon can also be used to reduce contamination from 
beam backgrounds such as beam-gas interactions, 
which form an important systematic uncertainty in the BES measurement.
Finally, the situation with respect to final-state radiation (FSR) is improved, with 
a clear geometrical separation between ISR, mostly located in the vicinity
of the collision axis, and FSR, which follows the directions of outgoing hadrons.
This can furthermore be verified in data by studying the forward-backward asymmetry 
produced by ISR/FSR interference.

\section{Conclusion}

The radiative return method offers promising prospects at \babar and other facilities. Competitive measurements can already be made
based an available datasets and preliminary results are expected in the coming months.


%

\begin{thebibliography}{99}
\bibitem{bnl} The Muon g-2 Collaboration, Phys. Rev. Lett. {\bf 89}:101804 (2002)
\bibitem{novo} The CMD-2 Collaboration, Phys. Lett. {\bf B527}:161-172 (2002)
\bibitem{kuehn1} S. Binner, J.H. Kuehn, K. Melnikov, Phys. Lett. {\bf B459} 279-287 (1999)
\bibitem{rodrigo2} German Rodrigo {\it et al.} Eur.Phys.J. {\bf C24}:71-82 (2002)
\bibitem{bes} The BES Collaboration, Phys. Rev. Lett. {\bf 88}:101802 (2002)
\end{thebibliography}
\end{document}